\documentclass[aps,prl,preprintnumbers,twocolumn,groupedaddress,showpacs]{revtex4}

\usepackage{latexsym}
\usepackage{amsmath}
\usepackage{amsfonts}
\usepackage{graphicx}

\newcommand{\bq}{\begin{eqnarray}}
\newcommand{\eq}{\end{eqnarray}}

\newcommand{\eps}{\varepsilon}

\begin{document}

\preprint{MZ-TH/08-22}
\title{\boldmath{NNLO corrections to 3-jet observables in electron-positron annihilation}}

\author{Stefan Weinzierl}
\affiliation{Institut f{\"u}r Physik, Universit\"at Mainz, D-55099 Mainz, Germany}

\date{July 21, 2008}

\begin{abstract}
I report on a numerical program, which can be used to calculate any infrared safe three-jet
observable in electron-positron annihilation to next-to-next-to-leading order in the strong 
coupling constant $\alpha_s$.
The results are compared to a recent calculation by another group.
Numerical differences in three colour factors are discussed and explained.
\end{abstract}


\maketitle

\section{Introduction}

Jet observables and event shapes in electron-positron annihilation can be used to extract
the value of the strong coupling constant $\alpha_s$ \cite{Biebel:2001dm,Kluth:2006bw,Bethke:2006ac}.
This applies in particular to three-jet observables, where the leading-order parton process
is proportional to $\alpha_s$.
In order to extract the numerical value from the LEP data, precise theoretical calculations
are necessary, calling for a next-to-next-to-leading order (NNLO) calculation.
Due to the large variety of interesting jet observables it is desirable not to perform
this calculation for a specific observable, but to set up a computer
program, which yields predictions for any infra-red safe observable relevant to the process
$e^+ e^- \rightarrow \mbox{3 jets}$.
Such a task requires the calculation of the relevant amplitudes up to two loops,
a method for the cancellation of infrared divergences and stable and efficient Monte
Carlo techniques.
For the process $e^+ e^- \rightarrow \mbox{2 jets}$ this was 
done in \cite{Anastasiou:2004qd,Gehrmann-DeRidder:2004tv,Weinzierl:2006ij,Weinzierl:2006yt}.
In this letter I report on a NNLO calculation for three-jet observables 
in electron-positron annihilation.
Recently another group published results for the NNLO corrections 
for three-jet observables \cite{GehrmannDeRidder:2008ug,GehrmannDeRidder:2007hr,GehrmannDeRidder:2007jk,GehrmannDeRidder:2007bj}.
In the calculation presented here the methods used are in many parts similar to the ones used 
in \cite{GehrmannDeRidder:2008ug,GehrmannDeRidder:2007hr,GehrmannDeRidder:2007jk,GehrmannDeRidder:2007bj}, although
I will show that in certain points there are important differences.
The authors of \cite{GehrmannDeRidder:2008ug,GehrmannDeRidder:2007hr,GehrmannDeRidder:2007jk,GehrmannDeRidder:2007bj} made
major contributions to the development of these methods \cite{Gehrmann-DeRidder:2003bm,Gehrmann-DeRidder:2004tv,Gehrmann-DeRidder:2005hi,Gehrmann-DeRidder:2005aw,Gehrmann-DeRidder:2005cm}.

The numerical results of the two calculations are compared.
The comparison is facilitated by splitting the NNLO correction term into individually gauge-invariant
contributions, such that each contribution is proportional to a specific colour factor.
For the NNLO corrections to $e^+ e^- \rightarrow \mbox{3 jets}$ there are six different colour factors.
In three colour factors the two calculations agree ($N_c^{-2}$, $N_f/N_c$, $N_f^2$).
They disagree in the remaining three colour factors ($N_c^2$, $N_c^0$, $N_f N_c$, ).
The numerical differences in these colour factors can be traced back to an incomplete cancellation
of soft-gluon singularities 
in the calculation 
of refs.~\cite{GehrmannDeRidder:2008ug,GehrmannDeRidder:2007hr,GehrmannDeRidder:2007jk,GehrmannDeRidder:2007bj}.
These singularities require additional subtraction terms, which are subtracted from the five-parton
configuration and added to the four-parton configuration. These subtraction terms have a structure
not present in \cite{GehrmannDeRidder:2007jk} and are related to soft gluons.
These terms occur generically in any NNLO calculation with three or more hard coloured partons.

\section{General set-up}

The perturbative expansion of any infrared-safe observable for the process
$e^+ e^- \rightarrow \mbox{3 jets}$
can be written up to NNLO as
\bq
{\cal O} & = & 
 \frac{\alpha_s}{2\pi} A_{\cal O}
 +
 \left( \frac{\alpha_s}{2\pi} \right)^2 B_{\cal O}
 +
 \left( \frac{\alpha_s}{2\pi} \right)^3 C_{\cal O}.
\eq
$A_{\cal O}$ gives the LO result, $B_{\cal O}$ the NLO correction and $C_{\cal O}$ the NNLO correction.
The coefficient $C_{\cal O}$ can be decomposed into colour pieces
\bq
 C_{\cal O} & = &
 \frac{1}{8} \left( N_c^2-1 \right)
 \left[ 
        N_c^2 C_{\cal O}^{lc}
      + C_{\cal O}^{sc}
      + \frac{1}{N_c^{2}} C_{\cal O}^{ssc}
 \right. \nonumber \\
 & & \left.
      + N_f N_c C_{\cal O}^{nf}
      + \frac{N_f}{N_c} C_{\cal O}^{nfsc}
      + N_f^2 C_{\cal O}^{nfnf}
 \right],
\eq
where $N_c$ denotes the number of colours and $N_f$ the number of light quark flavours.
In addition, there are singlet contributions, which arise from interference terms of amplitudes, where
the electro-weak boson couples to two different fermion lines.
These singlet contributions are expected to be numerically 
small \cite{Dixon:1997th,vanderBij:1988ac,Garland:2002ak}
and neglected in the present calculation.

The computation of the NNLO coefficient $C_{\cal O}$ requires the knowledge of the amplitudes
for the three-parton final state 
$e^+ e^- \rightarrow \bar{q} q g$ up to two-loops \cite{Garland:2002ak,Moch:2002hm},
the amplitudes of the four-parton final states
$e^+ e^- \rightarrow \bar{q} q g g$
and
$e^+ e^- \rightarrow \bar{q} q \bar{q} q$
up to one-loop \cite{Bern:1997ka,Bern:1997sc,Campbell:1997tv,Glover:1997eh}
and the five-parton final states
$e^+ e^- \rightarrow \bar{q} q g g g$
and
$e^+ e^- \rightarrow \bar{q} q \bar{q} q g$
at tree level \cite{Berends:1989yn,Hagiwara:1989pp}.
Taken separately, the three-, four- and five-parton contributions are all 
individually infrared divergent. Only the sum of them is finite.
However, the individual contributions live on different phase spaces, which prevents a naive Monte Carlo approach.
To render the individual contributions finite, several options for the cancellation of infrared divergences 
have been discussed, like phase space slicing \cite{Gehrmann-DeRidder:1998gf},
sector decomposition \cite{Binoth:2004jv,Heinrich:2006sw},
a method based on the optical theorem \cite{Anastasiou:2003gr}
or the subtraction method \cite{Kosower:2002su,Kosower:2003cz,Weinzierl:2003fx,Weinzierl:2003ra,Kilgore:2004ty,Frixione:2004is,Gehrmann-DeRidder:2003bm,Gehrmann-DeRidder:2004tv,Gehrmann-DeRidder:2005hi,Gehrmann-DeRidder:2005aw,Gehrmann-DeRidder:2005cm,Somogyi:2005xz,Somogyi:2006da,Somogyi:2006db,Catani:2007vq,Somogyi:2008fc,Aglietti:2008fe}.
In the present calculation I use the subtraction method 
with antenna subtraction terms \cite{Gehrmann-DeRidder:2005cm}.

\section{Cancellation of divergences}

To render the individual three-, four- and five-parton contributions finite, one adds and subtracts suitable chosen terms.
Schematically, we have
\bq
5\;\mbox{partons}: & &
 d\sigma_{5}^{(0)} 
 - d\alpha^{NLO}
 - d\alpha^{NNLO}
 + d\alpha^{iterated}
 \nonumber \\
 & &
 - d\alpha^{almost}
 - d\alpha^{soft},
 \nonumber \\
4\;\mbox{partons}: & &
 d\sigma_{4}^{(1)}
 + d\alpha^{NLO}
 - d\alpha^{loop}
 - d\alpha^{iterated}
 \nonumber \\
 & &
 - d\alpha^{product}
 + d\alpha^{almost}
 + d\alpha^{soft},
 \nonumber \\
3\;\mbox{partons}: & &
 d\sigma_{3}^{(2)}
 + d\alpha^{NNLO}
 + d\alpha^{loop}
 + d\alpha^{product}.
 \nonumber
\eq
Here, $d\sigma_{5}^{(0)}$, $d\sigma_{4}^{(1)}$ and $d\sigma_{3}^{(2)}$
are the contributions from the original amplitudes with five, four or three final state partons.
$d\alpha^{NLO}$ is the NLO subtraction term for four-jet observables, containing only
three parton tree-level antenna functions.
At NNLO there are several new subtraction terms required, each of them with a specific structure.
The term $d\alpha^{NNLO}$ contains the four-parton tree-level antenna functions.
The term $d\alpha^{loop}$ contains three-parton one-loop antenna functions together with tree-level
matrix elements and three-parton tree-level antenna functions together with one-loop matrix elements.
The remaining terms $d\alpha^{iterated}$, $d\alpha^{almost}$, $d\alpha^{product}$ and 
$d\alpha^{soft}$ all contain a product of two three-parton tree-level antenna functions. 
In $d\alpha^{iterated}$ and $d\alpha^{almost}$ one antenna function has five-parton kinematics, 
while the other antenna has four-parton kinematics. The former subtraction term is an approximation
to $d\alpha^{NLO}$, while the latter approximates $d\sigma_{5}^{(0)}$ in almost colour-correlated double
unresolved configurations.
In $d\alpha^{product}$ both antennas have four-parton kinematics.
The term $d\alpha^{soft}$ will be discussed below and is relevant only for the colour factors
$N_c^2$, $N_c^0$ and $N_f N_c$.

The subtraction terms without $d\alpha^{soft}$ correspond to the subtraction scheme of
ref.~\cite{GehrmannDeRidder:2007jk}.
For any subtraction scheme it is required, 
that in the three-parton channel the explicit divergences cancel, that the four-parton channel
is integrable over a single unresolved phase space and in addition that the explicit divergences cancel
and finally that in the five-parton channel the integrand is integrable over single and double
unresolved phase space regions.
It is easily checked that with the subtraction terms of ref.~\cite{GehrmannDeRidder:2007jk}
the explicit divergences in the three-parton cancel and I will focus in the following on the
four- and five-parton channels.

In the four-parton channel the combination $ d\sigma_{4}^{(1)} + d\alpha^{NLO}$
is free of explicit poles.
It has been noted in ref.~\cite{GehrmannDeRidder:2007jk} that the combination
$ d\alpha^{loop} + d\alpha^{iterated} + d\alpha^{product} - d\alpha^{almost}$
involves in the colour factors $N_c^2$ and $N_c^0$ poles of the form
\bq
\left| {\cal A}_3^{(0)}(1',2',j) \right|^2 X_3^0(1,i,2)
\frac{1}{\eps}
\left[ \ln \frac{s_{1'j} s_{j2'}}{s_{1'2'}} - \ln \frac{s_{1j} s_{j2}}{s_{12}} \right],
\nonumber
\eq
where $p_{1'}$ and $p_{2'}$ are the momenta obtained from $p_1$, $p_i$ and $p_2$ through a
$3 \rightarrow 2$ phase space map.
${\cal A}_3^{(0)}$ is the three-parton tree-level amplitude and $X_3^0(1,i,2)$ a three-parton
tree-level antenna function.
In ref.~\cite{GehrmannDeRidder:2007jk} it was claimed that these poles vanish
after the azimuthal integration over the unresolved phase space.
This claim is wrong.
In the centre-of-mass frame of $p_{1'}+p_{2'}$ with $p_{1'}$ and $p_1$ along the positive
$z$-axis, the relevant integral is
\bq
 I & = & 
\frac{1}{2\pi} \int\limits_0^{2\pi} d\phi
 \ln \left( \frac{(1+c_j)(1-c_2)}{2 (1-c_2c_j -s_2s_j\cos \phi)} \right),
\eq
where for $x=2,j$ we set
$c_x = \cos \theta_x$, $s_x = \sin \theta_x$ and $\theta_2$ and $\theta_j$ are the polar angles
of partons $2$ and $j$ in the chosen frame.
The integral equals
\bq
 I & = &
 \ln \left( \frac{1-c_2c_j+(c_j-c_2)}{1-c_2c_j+|c_j-c_2|} \right).
\eq
The integral is zero for $\theta_j < \theta_2$ but non-zero for $\theta_j > \theta_2$.
In ref.~\cite{GehrmannDeRidder:2007jk} it was claimed that the integral vanishes in both cases.
As a consequence of the non-zero value for $\theta_j > \theta_2$
the explicit poles do not cancel in the combination
$ d\alpha^{loop} + d\alpha^{iterated} + d\alpha^{product} - d\alpha^{almost}$.
The same situation occurs also in the colour factor $N_f N_c$.

These poles have a counter-part in the five-parton channel. Setting $d\alpha^{soft}$ to zero
\begin{figure}[ht]
\begin{center}
\includegraphics[bb= 125 460 490 710,width=0.5\textwidth]{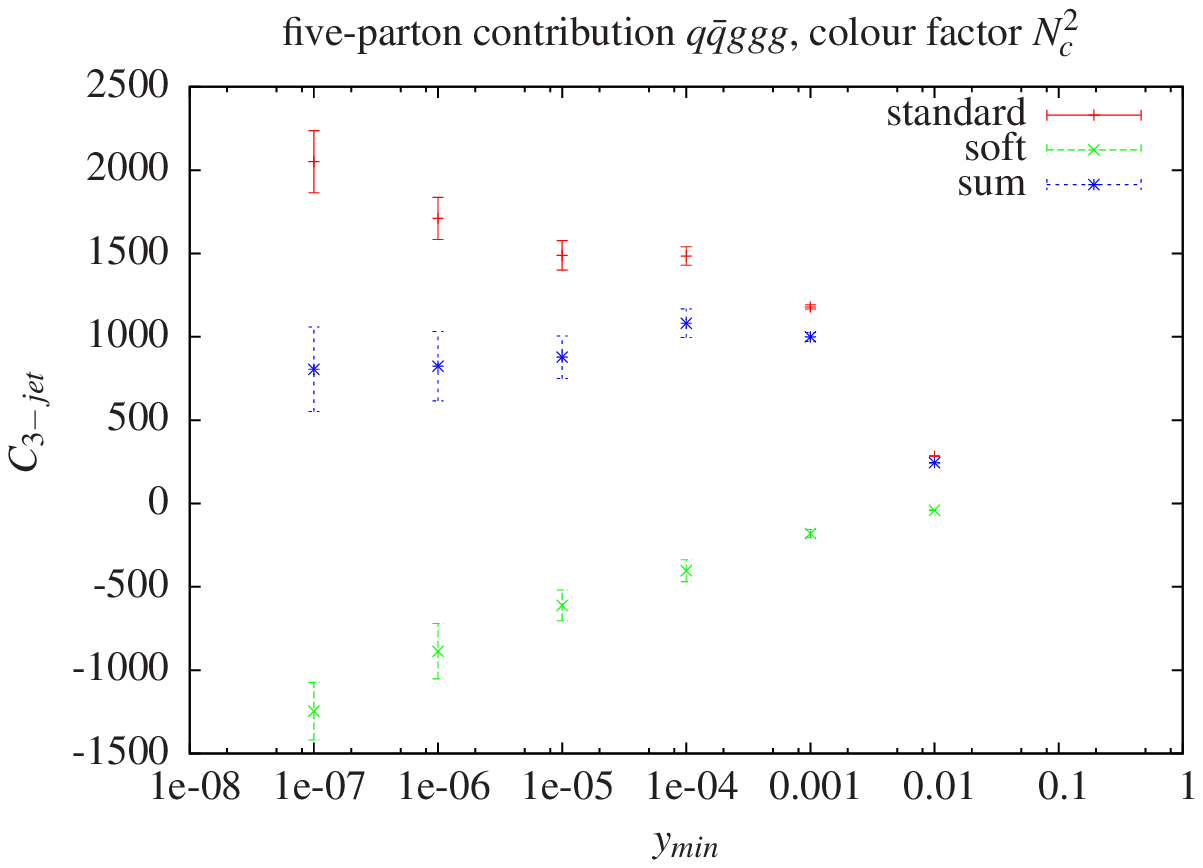}
\includegraphics[bb= 125 460 490 710,width=0.5\textwidth]{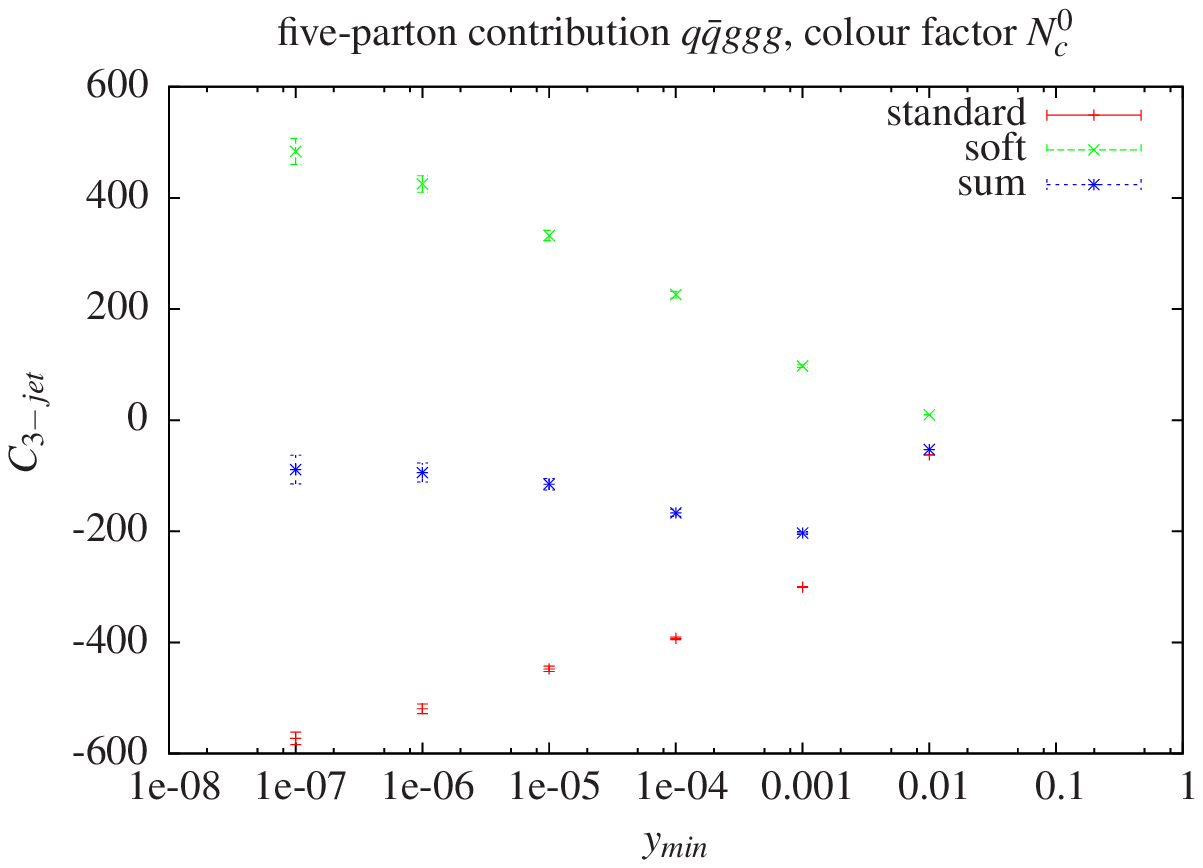}
\includegraphics[bb= 125 460 490 710,width=0.5\textwidth]{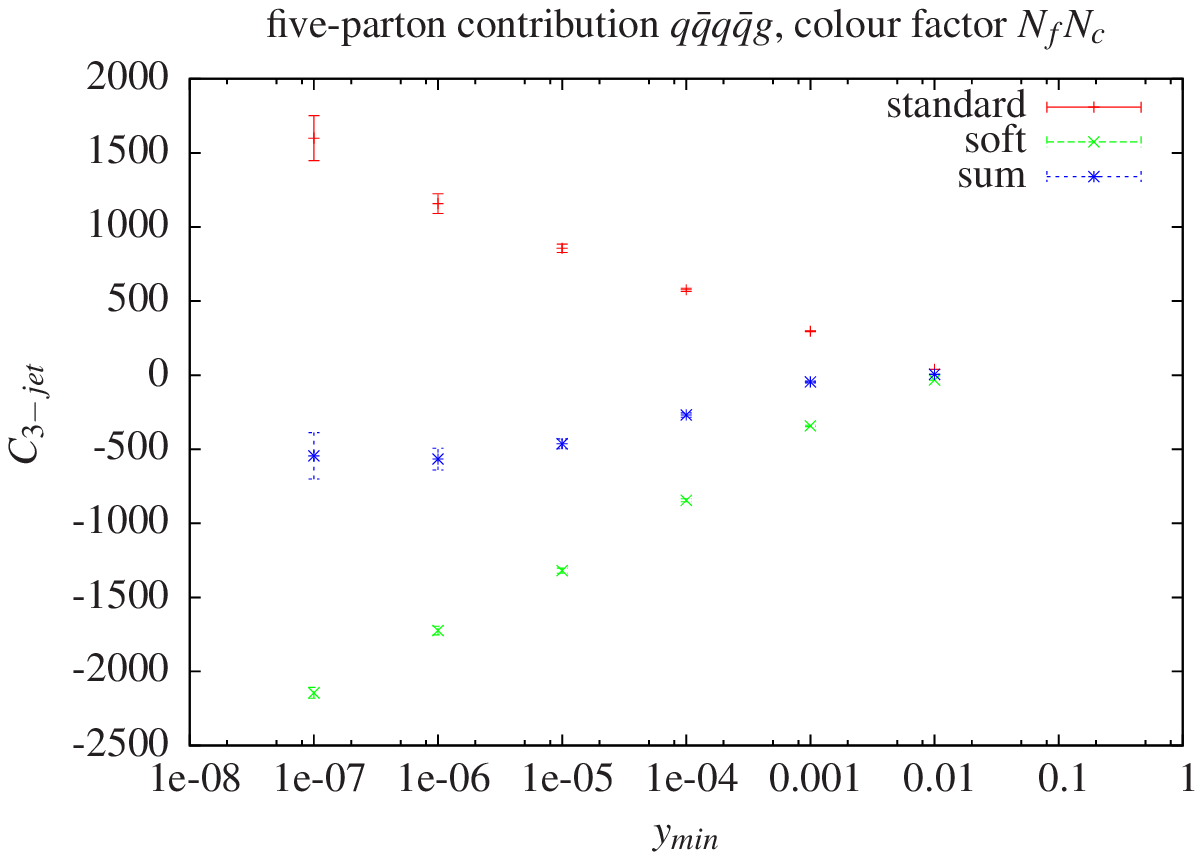}
\end{center}
\caption{Dependence of the five-parton contribution on the slicing parameter $y_{min}$
for the Durham jet cross section with $y_{cut}=0.01$ in the colour factors $N_c^2$, $N_c^0$ and $N_f N_c$.
``Standard'' denotes the combination 
$d\sigma_{5}^{(0)} - d\alpha^{NLO} - d\alpha^{NNLO} + d\alpha^{iterated} - d\alpha^{almost}$,
``soft'' the contribution from $d\alpha^{soft}$. In addition the
sum of the two terms is shown.
For small values of $y_{min}$ the sum is independent of $y_{min}$.
}
\label{fig1}
\end{figure}
and using a slicing approach one observes in the colour factors $N_c^2$, $N_c^0$ and $N_f N_c$ 
a logarithmic dependence on the slicing parameter $y_{min}=s_{min}/Q^2$.
This is shown in fig.~\ref{fig1}.

These singularities require an additional subtraction term and that is where the present calculation
differs from the one of ref.~\cite{GehrmannDeRidder:2008ug,GehrmannDeRidder:2007hr,GehrmannDeRidder:2007jk,GehrmannDeRidder:2007bj}.
$d\alpha^{soft}$ is a subtraction term related to soft gluons which ensures that the poles in the four-parton
configuration vanish after integration over the azimuthal angle and which renders
the five-parton configuration independent of $y_{min}$.
The term $d\alpha^{soft}$ for the four-parton configuration can be taken of the form
\bq
\label{soft_subtr}
\lefteqn{
\left| {\cal A}_3^{(0)}(1',2',j) \right|^2 
 X_3^0(1,i,2)
 \theta\left( \frac{2p_1p_j}{2p_{1i2}p_j} - \frac{2p_1p_2}{2p_{1i2}p_2} \right)
} & & 
 \nonumber \\
 & &
\left[ 
       {\cal S}_3^0\left( s_{1j} \right)
     - {\cal S}_3^0\left( s_{12} \right)
     - {\cal S}_3^0\left( s_{\hat{2}j} \right)
     + {\cal S}_3^0\left( s_{\hat{2}2} \right)
\right],
\eq
where ${\cal S}_3^0$ is the integrated soft antenna function and $p_{\hat{2}}$ is given by
\bq
 p_{\hat{2}} & = & 
 p_2 + p_i - \frac{s_{2i}}{s_{12}+s_{1i}} p_1.
\eq
I also used the short-hand notation $p_{1i2}=p_1+p_i+p_2$.
The $\theta$-function enforces $\theta_j > \theta_2$
in the specific frame introduced above.
$d\alpha^{soft}$ for the five-parton configuration is obtained by lifting eq.~\ref{soft_subtr} to the
five parton phase space.
Fig.~\ref{fig1} shows that the sum of all contributions in the five-parton channel is now independent
of $y_{min}$.
I have checked that in the four-parton channel the explicit poles cancel after integration over the unresolved phase space.

\section{Numerical results}

\begin{figure}[ht]
\begin{center}
\includegraphics[bb= 125 460 490 710,width=0.5\textwidth]{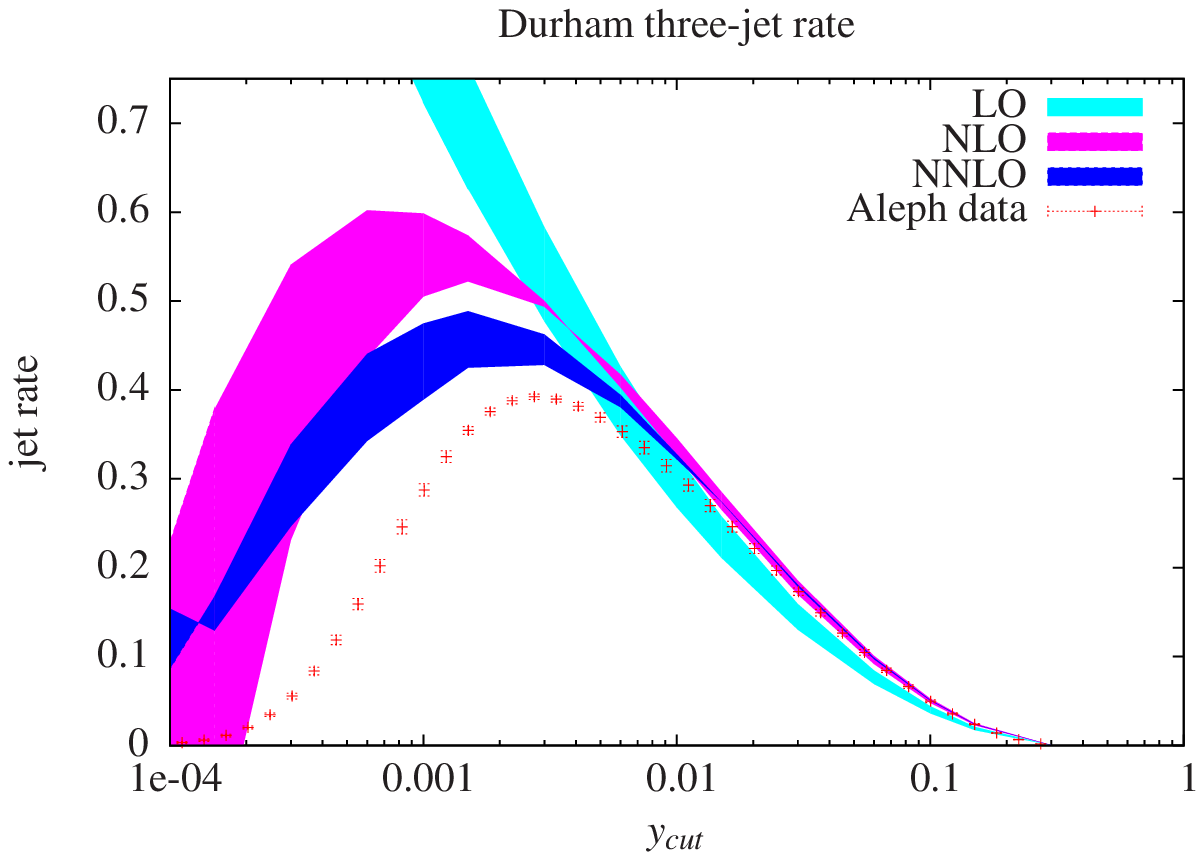}
\end{center}
\caption{The scale variation of the three jet rate with the Durham jet algorithm at $\sqrt{Q^2}=m_Z$ with $\alpha_s(m_Z)=0.118$.
The bands give the range for the theoretical prediction obtained from varying the renormalisation scale
from $\mu=m_Z/2$ to $\mu=2 m_Z$.
}
\label{fig2}
\end{figure}

The numerical program is build on an existing NLO program for 
$e^+ e^- \rightarrow \mbox{4 jets}$ \cite{Weinzierl:1999yf}.
I consider the three-jet cross section, where the jets are defined 
by the Durham jet algorithm \cite{Stirling:1991ds}.
The recombination prescription is given by the E-scheme.
I take the centre of mass energy to be $\sqrt{Q^2} = m_Z$.
The three-jet cross section is expanded as
\bq
\lefteqn{
\sigma_{3-jet} = 
} & & \nonumber \\
 & &
 \sigma_0 \left[  
 \frac{\alpha_s}{2\pi} A_{3-jet}
 +
 \left( \frac{\alpha_s}{2\pi} \right)^2 B_{3-jet}
 +
 \left( \frac{\alpha_s}{2\pi} \right)^3 C_{3-jet}
 \right],
 \nonumber 
\eq
where $\sigma_0$ is the LO cross section for $e^+ e^- \rightarrow \mbox{hadrons}$.
The coefficients $A_{3-jet}$, $B_{3-jet}$ and $C_{3-jet}$ are given for 
the renormalisation scale $\mu^2 = Q^2$ and various values of the jet defining parameter $y_{cut}$ in table~\ref{tableABC}.
\begin{table}
\begin{center}
\begin{tabular}{|l|lll|}
\hline
 $y_{cut}$ & $A_{3-jet}$ & $B_{3-jet}$ & $C_{3-jet}$ \\
\hline
 $0.3$ &    $0.02$ & $0.13$             & $-6 \pm 3$ \\
 $0.1$ &    $2.12$ & $34.3$             & $(2.0 \pm 0.2 ) \cdot 10^2 $ \\
 $0.03$ &   $7.63$ & $113.8$            & $(6.7 \pm 0.6 ) \cdot 10^2 $ \\
 $0.01$ &   $15.7$ & $152.6$            & $(-1.2 \pm 0.2) \cdot 10^3$ \\
 $0.003$ &  $27.9$ & $-6.5$             & $(-8.1 \pm 0.5) \cdot 10^3$ \\
 $0.001$ &  $42.4$ & $-562$             & $(-21 \pm 1 ) \cdot 10^3$ \\
 $0.0003$ & $61.8$ & $-1.97 \cdot 10^3$ & $(-25 \pm 3) \cdot 10^3$ \\
 $0.0001$ & $82.9$ & $-4.36 \cdot 10^3$ & $(7 \pm 5) \cdot 10^3$ \\
\hline
\end{tabular}
\caption{\label{tableABC}
The LO coefficient $A_{3-jet}$, the NLO coefficient $B_{3-jet}$ and the NNLO coefficient $C_{3-jet}$
for the three jet cross section with the Durham jet algorithm and various values of $y_{cut}$.
}
\end{center}
\end{table}
The errors of $C_{3-jet}$ are from the Monte Carlo integration.
For selected values of $y_{cut}$ the contribution from the individual colour factors to the NNLO coefficient
$C_{3-jet}$ is shown in table~\ref{table_colour}.
\begin{table}
\begin{center}
\begin{tabular}{|l|ll|}
\hline
 $y_{cut}$ & $N_c^2$ & $N_f N_c$ \\
\hline
 $0.1$ &   $(1.06 \pm 0.02) \cdot 10^3$ & $(-9.80 \pm 0.06) \cdot 10^2$ \\
 $0.01$ &  $(4.6  \pm 0.2)\cdot 10^3$ & $(-8.11 \pm 0.03) \cdot 10^3$ \\
 $0.001$ & $(-29 \pm 1 )\cdot 10^3$     & $(-2.7 \pm 0.2) \cdot 10^3$   \\
\hline
\hline
 $y_{cut}$ & $N_c^0$ & $N_f/N_c$ \\
\hline
 $0.1$ &   $-35 \pm 1$                  & $21.9 \pm 0.3$ \\
 $0.01$ &  $(9.7 \pm 0.1)\cdot 10^2$    & $(-2.66 \pm 0.02) \cdot 10^2$ \\
 $0.001$ & $(7.09 \pm 0.08) \cdot 10^3$ & $(-4.43 \pm 0.01) \cdot 10^3$ \\
\hline
\hline
 $y_{cut}$ & $N_c^{-2}$ & $N_f^2$ \\
\hline
 $0.1$ &   $-0.49 \pm 0.03$             & $(1.336 \pm 0.003) \cdot 10^2$ \\
 $0.01$ &  $0.25 \pm 0.15$              & $(1.646 \pm 0.002) \cdot 10^3$ \\
 $0.001$ & $(3.38 \pm 0.01) \cdot 10^2$ & $(7.41 \pm 0.01) \cdot 10^3$   \\
\hline
\end{tabular}
\caption{\label{table_colour}
The contributions from the individual colour factors to the NNLO coefficient $C_{3-jet}$.
}
\end{center}
\end{table}
Finally, fig.~\ref{fig2} shows the scale variation of the jet rate defined
by
\bq
 \frac{\sigma_{3-jet}}{\sigma_{tot}}
 & = &
 \frac{\alpha_s}{2\pi} \bar{A}_{3-jet}
 +
 \left( \frac{\alpha_s}{2\pi} \right)^2 \bar{B}_{3-jet}
 +
 \left( \frac{\alpha_s}{2\pi} \right)^3 \bar{C}_{3-jet},
 \nonumber
\eq
where
\bq
& &
 \bar{A}_{3-jet} = A_{3-jet},
 \;\;\;
 \bar{B}_{3-jet} = B_{3-jet} - A_{3-jet} A_{tot},
 \nonumber \\
& &
 \bar{C}_{3-jet} = C_{3-jet} - B_{3-jet} A_{tot} - A_{3-jet} \left( B_{tot} - A_{tot}^2 \right)
 \nonumber 
\eq
and $A_{tot} = 2$,
\bq
\lefteqn{
 B_{tot} = } & &   
\nonumber \\
 & & \frac{N_c^2-1}{8 N_c} \left[ \left( \frac{243}{4} - 44 \zeta_3 \right) N_c + \frac{3}{4 N_c} 
                             + \left( 8 \zeta_3 - 11 \right) N_f \right]. 
 \nonumber 
\eq
The renormalisation scale is varied
from $\mu=m_Z/2$ to $\mu=2 m_Z$.
In this plot the experimental measured values are also shown \cite{Heister:2003aj}.
For values below $y_{cut}=0.001$ the results of ref.~\cite{GehrmannDeRidder:2008ug} 
differ significantly from the ones presented here.

\section{Conclusions}

In this letter I reported on the NNLO calculation for three-jet
observable in electron-positron annihilation.
Particular attention was paid to the cancellation of infrared singularities.
I presented numerical results for the Durham three-jet cross section.

\subsection*{Acknowledgements}

I would like to thank Th.~Gehrmann for useful discussions and for
providing me with the results of \cite{GehrmannDeRidder:2008ug} for the individual colour factors.


\bibliography{/home/stefanw/notes/biblio}

\end{document}